\documentclass[preprint,aps,showpacs]{revtex4}

\usepackage{graphicx}% Include figure files
\usepackage{dcolumn}% Align table columns on decimal point
\usepackage{bm}% bold math

\begin{document}
\title{Phenomenological model of propagation of the elastic
waves in a fluid-saturated porous solid with 
non-zero boundary slip velocity}
\author{David Tsiklauri}
%\email{tsikd@astro.warwick.ac.uk}
%\homepage{http://www.astro.warwick.ac.uk/~tsikd}
\affiliation{Physics Department, University of Warwick, Coventry, CV4 7AL, 
United Kingdom}

\begin{abstract}
\vskip 2cm
It is known that a boundary slip velocity
starts to play important role when
the length scale over
which the fluid velocity changes approaches the slip length,
i.e. when  the fluid is highly confined, 
for example, fluid flow through porous rock or 
blood vessel capillaries.
Zhu \& Granick [Phys. Rev. Lett. {\bf 87}, 096105 (2001)] 
have recently experimentally established
existence  of a boundary slip in a Newtonian liquid.
They reported typical values of the slip length of the order of  few
micro-meters. In this light,
the  effect of introduction of the boundary slip 
into the theory of propagation 
of elastic waves in a fluid-saturated
porous medium  formulated by Biot 
[J. Acoust. Soc. Am., {\bf 28}, 179-191 (1956)] is
investigated.
Namely, the
effect of introduction of boundary slip upon
the function $F(\kappa)$
 that measures the deviation from
Poiseuille flow friction as a function of frequency 
parameter $\kappa$ is studied. By postulating {\it phenomenological}
dependence  of the slip velocity upon frequency,
notable  deviations  in the
domain of intermediate frequencies, in the 
behavior of $F(\kappa)$ are introduced with the incorporation
of the boundary slip into the model.
It is known that $F(\kappa)$ crucially enters Biot's equations,
which describe dynamics of fluid-saturated porous solid.
Thus, consequences of the non-zero
boundary slip by calculating
 the  phase velocities
and attenuation coefficients of the both rotational 
and dilatational waves with the variation of frequency are investigated.
The new model should allow to fit
the experimental seismic data in circumstances
when Biot's theory fails, as the introduction of phenomenological
dependence of the slip velocity upon frequency, which
is based on robust physical arguments, adds an additional
degree of freedom to the model.
If fact, it predicts {\it higher than the Biot's theory 
values of attenuation
coefficients of the both rotational 
and dilatational waves} in the intermediate frequency domain, 
which is in qualitative agreement with 
the experimental data.
Therefore, the introduction of the boundary slip yields three-fold benefits:
(A) Better agreement of theory with an experimental
data since the parametric space of the model is larger
(includes effects of boundary slip);
(B) Possibility to identify types of porous medium and physical
situations where boundary slip is important;
(C) Constrain model parameters that are related to the
boundary slip.
\end{abstract}
\date{\today}
\pacs{43.20.Jr; 47.27.Lx; 47.55.Mh; 47.60.+i; 62.65.+k}
\maketitle

\section{Introduction}

It has been a common practice in the
 fluid dynamics to assume
that when a fluid flows over an interface with a solid,
the fluid molecules adjacent to the solid have zero relative
velocity with the respect to solid. So far, this widely used assumption, 
known as "no-slip boundary condition", has been  successfully 
applied  to the theoretical modeling of almost all macroscopic experiments.
As relevantly noticed by Craig  et al. \cite{craig}, the success of
this assumption does not reflect its accuracy, but rather insensitivity
of the experiment to a partial-slip boundary condition.
It is known that the boundary slip becomes important only 
when the length scale over
which the fluid velocity changes approaches the slip length, 
that is the distance behind the interface at which the fluid
velocity extrapolates to zero, i.e. when  the fluid becomes highly confined, 
e.g., blood flow through capillaries 
or fluid flow through natural porous rock. 
Recently, authors of Refs. \cite{craig,zg} presented
a convincing experimental evidence of a boundary
slip in a Newtonian liquid. 
They performed direct measurements of the hydrodynamic
drainage forces, which show a clear evidence of boundary slip.
Also, they found that the boundary slip is a function of the fluid viscosity
and the shear rate. These results have important implications
for the blood dynamics in capillaries, the permeability of 
porous media, and lubrication of nano-machines.
For example, results of Craig et al. suggest that red
blood cells squeeze through narrow capillary walls more easily
and induce less shear stress on capillary walls due to the boundary
slip. Also, in oil production industry,
the residual oil is difficult to produce due to its naturally 
low mobility. Thus, the enhanced oil recovery operations are 
used to increase production.  It has been experimentally proven
and theoretically validated
that there is a substantial increase in the net fluid flow 
through a porous medium if the latter is treated with elastic waves
\cite{ore1,ore2,ore3,ore4}. We may conjecture that the elastic waves via the
pore wall vibration cause boundary slip of the residual oil
droplets, which likewise red blood cells, squeeze through
pores with less resistance, effectively increasing
permeability of the porous medium.

A quantitative theory of propagation 
of elastic waves in a fluid-saturated
porous solid has been formulated in the classic paper by Biot \cite{biot}.
After its appearance, this theory has seen numerous modifications and
generalizations. One of the major findings of Biot's work was 
that there was
a breakdown in Poiseuille flow above a certain characteristic frequency
specific to this fluid-saturated porous material. 
Biot theoretically studied this breakdown by
considering the flow of a viscous fluid in a tube with
longitudinally oscillating walls under an oscillatory pressure
gradient. 
Biot's theory can be used to describe interaction of fluid-saturated 
porous solid with sound for a classic Newtonian fluid assuming
no-slip boundary condition at the pore walls holds.
However, in the light of recent experimental results of 
Ref. \cite{zg}, revision of the classic
theory is needed in order to investigate novelties
bought about by the boundary slip. 

Biot's theory
has been a successful tool for interpreting the experimental
data for decades, but there are circumstances when it fails.
Gist \cite{gist} performed 
ultrasonic velocity and attenuation measurements in sandstones with 
a variety of saturating fluids and compared the data 
with the predictions of the
Biot's theory.
{\it Velocity data show systematic deviations from Biot theory
as a function of pore fluid viscosity. Ultrasonic attenuation is much 
larger than the Biot prediction}
and for several sandstones is nearly constant for a three decade variation 
in viscosity. This behavior contrasts with synthetic porous media such as 
sintered glass
beads, where Biot theory provides accurate predictions of both velocity and 
attenuation. The similar claim has been also made by earlier 
experimental studies e.g. Ref.~\cite{mochizuki}.
Even Biot himself acknowledged \cite{b62} that the model of a
purely elastic solid matrix saturated with a viscous
fluid is rather idealistic and there are rare occasions
where it gives satisfactory agreement with the experiment.
Thus, he generalized theory applying the correspondence
principle, and incorporated the solid and fluid attenuation, and
dissipation due to non-connected pores. In this work
Biot introduced the viscodynamic operator, which
is a sum of the viscosity and inertial terms, and it
accounts for the dynamic properties of the fluid motions
in pores valid for the both low and high frequencies.
The theory can model other, more complex, mechanisms
such as thermoelastic dissipation, which produce a
relaxation spectrum, or local flow mechanisms as the
'squirt' flow \cite{mwk86}, \cite{dnn94}.
A time domain formulation based on memory variables has been
also successfully introduced \cite{c93}, \cite{cq96}.
In order to narrow down the source of discrepancy between 
Biot theory and experiment, Gist \cite{gist} tried to modify
natural sandstone by curing a residual saturation of epoxy in the pore
space, filling small pores and micro-cracks. 
This altered rock has a significantly reduced attenuation, 
demonstrating the dominance of small pores in controlling
ultrasonic attenuation. Then he suggested that two non-Biot 
viscous attenuation mechanisms are needed: local flow in micro-cracks 
along grain contacts, and
attenuation from pore-wall surface roughness.
The model model based on these two mechanisms 
seemed to succeed in improvement of fitting the data.
However, this by no means does not exclude other possibilities,
and all plausible effects that could explain the discrepancy
should be explored.
At low frequencies, the fluid flow is of Poiseuille type
and the inertia effects are obviously negligible in comparison with
the effects of the viscosity. At high frequencies, however,
these effects are confined to a thin boundary layer in the
vicinity of the pore walls and the inertia forces are dominant.
Therefore, {\it we conjecture that in such situations, apart from
other yet unknown effects, non-zero boundary slip effect maybe
responsible for the deviations between the theory and
experiment}. This justifies our aim to formulate
a model that would account for the boundary slip.

In the section II we formulate theoretical basis of our model and
in section III we present our numerical results. In the
section IV we conclude with the discussion of the results.

\section{The model}

In  our model we study  a Newtonian fluid flowing in
a cylindrical tube, which mimics a natural pore,
whose walls are oscillating longitudinally
and the fluid is subject to an oscillatory pressure gradient.
We give analytical solutions of the problem in the frequency domain.

The governing equation of the problem is
 the linearized momentum equation of an incompressive fluid
$$
\rho {{\partial \vec v}\over{\partial t}} = - \nabla p + \mu \nabla^2 \vec v.
\eqno(1)
$$ 
Here, $\vec v$, $p$, $\rho$ denote velocity, pressure and mass density
of the fluid, while $\mu$ is the viscosity coefficient.

Now, let $u$ be a velocity of the wall of the tube which oscillates in time 
as $e^{- i \omega t}$. 
The flow of fluid in a cylindrical tube with longitudinally
oscillating walls can be described by a singe component of
the velocity, namely, its $z$-component $v_z$ ($z$ axis is along 
the centerline of the tube). We use cylindrical coordinate
system $(r,\phi,z)$ in treatment of the problem.
We introduce the relative velocity $U_1$ as
$U_1=v_z-u$. Thus, assuming that all physical quantities vary in time as
$e^{- i \omega t}$, we arrive at the following master equation for $U_1$
$$
\nabla^2 U_1 +
{{ i \omega}\over{\nu}} U_1=
 -{{X}\over{\nu}}. 
\eqno(2)
$$
Here, we have introduced the following notations:
$$
\rho X= -(\nabla p + \rho{{\partial u}\over{\partial t}} ),
$$
which is a sum of the applied pressure gradient and force exerted
on the fluid from the oscillating wall of the tube and,
$\nu$, which is $\nu= \mu / \rho$.

The solution of Eq.(2) can be found to be \cite{biot}
$$
U_1 (r)= - {{X}\over{i \omega}} + C J_0(\beta r),
$$
where $J_0$ is the Bessel function and 
$\beta = \sqrt{ i \omega/{\nu}}$.

Assuming that the slip
velocity is $U_1(a)=U_s$  at the 
wall of the tube, where $a$ is its radius, we  obtain
$$
U_1(r)= - {{X}\over{i \omega}} 
\left[1 - (1+\bar U_s){{J_0(\beta r)}\over{J_0(\beta a)}} \right].
\eqno(3)
$$
Here, 
$$
\bar U_s \equiv U_s {{i \omega}\over{X}}
= U_s {{\nu}\over{a^2X}} (\beta a)^2.
$$

Defining the cross-section averaged velocity as 
$$
\bar U_1 ={{2}\over{a^2}} \int_0^a U_1(r) r d r,
$$
we obtain
$$
\bar U_1=-{{Xa^2}\over{ \nu }} {{1}\over{(\beta a)^2}} 
\left[1 - {{2 (1+\bar U_s) 
J_1(\beta a)}\over{(\beta a) J_0(\beta a)}} \right].
\eqno(4)
$$

Following work of Biot \cite{biot} we calculate the stress at the
wall $\tau$,
$$
\tau = -{\mu}
\left({{\partial U_1(r)}\over{\partial r}}\right)_{r=a}
= {{\mu \beta X}\over{i \omega}}
\left(1+\bar U_s \right){{J_1(\beta a)}\over{J_0(\beta a)}}.
\eqno(5)
$$

The total friction force is $2 \pi a \tau$. Following
Biot we calculate the ratio of total friction
force to the average velocity, i.e.
$$
{{2 \pi a \mu}\over{\bar U_1}}=
- 2 \pi \mu (\beta a)
\left(1+\bar U_s \right) {{J_1(\beta a)} \over{J_0(\beta a)}}
$$
$$
\times \left[1 - {{2 \left(1+\bar U_s \right) J_1(\beta a)}\over
{(\beta a) J_0(\beta a)}} \right]^{-1}.
\eqno(6)
$$
Simple analysis reveals that (assuming $\bar U_s \to 0$ as $\omega \to 0$,
see discussion below)
$$
\lim_{\omega \to 0}
{{2 \pi a \tau}\over{\bar U_1}}={8 \pi \mu},
$$
which corresponds to the limiting case of
Poiseuille flow. 
Following Biot \cite{biot}, we also introduce a function
$F(\kappa)$ with $\kappa$ being frequency parameter,
$\kappa= a \sqrt{\omega / \nu}$, in the following
manner
$$
{{2 \pi a \tau}\over{\bar U_1}}=
8 \pi \mu F(\kappa),
$$
thus,
$$
F(\kappa)=-{{1}\over{4}}
 \kappa \sqrt{ i }
\left(1+\bar U_s \right){{J_1(\kappa \sqrt{i})}\over{
 J_0(\kappa \sqrt{i})}}
$$
$$
\times \left[1 - \left(1+\bar U_s \right)
{{2 J_1(\kappa \sqrt{i})}\over
{\kappa \sqrt{i}
J_0(\kappa \sqrt{i})}}
\right]^{-1}.
\eqno(7)
$$
Note, that $F(\kappa)$  measures the deviation from
Poiseuille flow friction as a function of frequency parameter $\kappa$.
The  Biot's expression for $F(\kappa)$
in the no boundary slip regime can be easily recovered from 
Eq.(7) by putting $\bar U_s \to 0$ for all $\kappa$'s.

So far, we did not specify $\bar U_s$,  however there are
certain physical constraints it should satisfy:

(A) Authors of  Ref.~\cite{craig} demonstrated that
 the slip length (which, in fact, is proportional to the
slip velocity) is a function of the approach rate and
they showed why several previous careful measurement of confined
liquids have not observed evidence for boundary slip.
Under the low approach rates employed in previous measurements
slip length is zero and no-slip boundary condition is
applicable. 
Experiments reported in Ref.~\cite{craig} were performed when
half-sphere approached a plane with liquid placed
in-between at different approach rates.
However, we should clearly realize what
term "approach rate" means in the context of Biot's
theory: Biot investigated fluid flow in the
cylindrical tube whose walls are harmonically
oscillating in the longitudinal direction as
$x(t)=A e^{-i \omega t}$, therefore if similar to
Ref.\cite{craig} experiment would be done for the
oscillation tube, the "approach rate" would be
the amplitude of $\dot x(t)$, i.e. $-i \omega A$.
Thus, when $\omega \to 0$, $\bar U_s$ should also tend to
zero.

(B) At high frequencies the viscous effects are
negligible compared to the inertial effects.
Thus, fluid at high frequencies behaves as an ideal (non-viscous)
pore fluid, which allows for an arbitrary slip.
Therefore, when $\omega \to \infty$, $\bar U_s$ should  tend to
zero. When we examine Fig.~6 from Ref.\cite{craig} and
Fig.~3 (bottom panel) from Ref.\cite{zg}
closely
we should understand that the authors measured ascending
part where slip length increases with the increase of
the approach (flow) rate. Since, the viscous effects should be
negligible compared to the inertial effects for
large approach rates, there must be also a descending
part -- clearly slip length (or the same as slip velocity)
cannot grow infinitely as approach rate increases.
Viscous effects should give in to the inertial effects
and fluid should behave an ideal fluid allowing for an
arbitrary slip.

Based upon the above two physical arguments we conclude that
$\bar U_s$ should be a smooth function which tends to
zero when $\omega$ (or the same as $\kappa$) tends to both
zero and $\infty$. Therefore, we postulate following
{\it phenomenological}
expression for $\bar U_s$
$$
Re [\bar U_s(\kappa)]=Im [\bar U_s(\kappa)] = \xi {{B \kappa^4}\over{(A+\kappa^2)^5}}
$$
This function has a maximum at $\kappa_*=\sqrt{2A/3}$. By fixing
$B$ at $5^5 A^3/(4 \times 3^3)$ we force it to achieve its maximum equal
to unity at $\kappa=\kappa_*$ (when $\xi=1$). 
We introduced $\xi$ (usually $0 \leq \xi \leq 1$) 
as a sort of "weight" of the boundary slip
effect into the solution. Putting, $\xi=0$ this way would
easily allow us to recover no-slip case, while the increase of
$\xi$ we would be able to trace effect of non-zero slip onto
the all physical quantities.
We plot this function in Fig.~1 for $A=25$ and $\xi=1$.

It is worthwhile to mention that, at first glance, it seems
that non-zero boundary slip, 
which appears at "high approach rates",
should be attributed to 
the non-linear effects. However,
thorough interpretation of the experimental results of Ref.~\cite{craig},
in the context of the oscillatory tube (see points (A) and (B) above)
allows us to conclude that the non-zero boundary slip
can be also incorporated into the {\it linear} Biot's
theory.

Next,  we plot both $F_r(\kappa)=Re[F (\kappa)]$
and $F_i(\kappa)=-Im[F (\kappa)]$ in Fig.~2 for the three cases: when
there is no boundary slip ($\xi=0$), and $\xi=0.05, \; 0.1$.
We gather from the plot that the $\xi=0$ is identical
to Fig.~4 from Ref.\cite{biot} as it should be. However,
we also notice a noticeable difference from the classic case
when $\xi$ is non-zero in the {\it intermediate} frequencies
domain. Of course, according to our definition of the
phenomenological form of $\bar U_s(\kappa)$, even for
non-zero $\xi$ (non-zero boundary slip), 
when $\kappa \to 0$ and  $\kappa \to \infty$, asymptotically
$F(\kappa)$ behaves as classic Biot's solution \cite{biot}, i.e.
$$
\lim_{\kappa \to 0}F_r(\kappa)= 1,
\eqno(8)
$$
and
$$
\lim_{\kappa \to 0}F_i(\kappa)= 0,
\eqno(9)
$$

$$
\lim_{\kappa \to \infty}F(\kappa)=
{{\kappa}\over{4}} \sqrt{i}={{\kappa}\over{4}}\left(
{{1+i}\over{\sqrt{2}}}\right).
\eqno(10)
$$
Since in our phenomenological model we allow for
the deviations in the intermediate frequency domain,
it is easy to foresee that
these will have an impact on all of the predictions
of the Biot's theory precisely in that frequency range. 
Namely, all observable quantities
predicted by the Biot's theory, such as phase velocities
and attenuation coefficients of the both rotational and dilatational
waves will be affected by the introduction
of boundary slip into the model in the intermediate frequency range.

Biot \cite{biot} showed  that the general equations  
which govern propagation
of rotational and dilatational high-frequency waves in a 
fluid-saturated porous medium are the same as in the low-frequency 
range provided the viscosity is replaced by its
effective value as a function of frequency.
In practice, it means replacing the resistance
coefficient $b$ by $bF(\kappa)$.

The equations describing dynamics of the rotational waves are \cite{biot}
$$
{{\partial^2}\over{\partial t^2}}(\rho_{11}\vec \omega + \rho_{12}\vec
\Omega)
+b F(\kappa){{\partial}\over{\partial t}}(\vec \omega - \vec \Omega)=
N \nabla^2 \vec \omega,
\eqno(11)
$$
$$
{{\partial^2}\over{\partial t^2}}(\rho_{12}\vec \omega + \rho_{22}\vec
\Omega)
- b F(\kappa){{\partial}\over{\partial t}}(\vec \omega - \vec \Omega)=0,
\eqno(12)
$$
where, $\rho_{11},\rho_{12}$ and $\rho_{22}$ are mass density
parameters for the solid and fluid and their inertia coupling; 
$\vec \omega= \mathrm{curl}\; \vec u$ and $\vec \Omega =\mathrm{curl}\; 
\vec U$  describe 
rotations of solid and fluid with $\vec u$ and $\vec U$ 
being their displacement 
vectors, while
the rigidity of the solid is represented by the modulus $N$.
Substitution of a plane rotational
wave of the form
$$
\omega=C_1 e^{i(l x + \chi t)}, \;\;\;
\Omega=C_2 e^{i(l x + \chi t)},
\eqno(13)
$$
into Eqs.(11) and (12) allows us to obtain a characteristic equation
$$
{{Nl^2}\over{\rho a^2}}=E_r -i E_i,
\eqno(14)
$$
where $l$ is wavenumber, $\chi=2 \pi f$ is  wave cyclic frequency,
 $\rho=\rho_{11}+2 \rho_{12}+\rho_{22}$ is the mass
density of the bulk material and $a$ is a pore radius.

The real and imaginary parts of  Eq.(14) can be written
as
$$
E_r={{(\gamma_{11}\gamma_{22}-\gamma_{12}^2)(\gamma_{22}
+\epsilon_2)+\gamma_{22}\epsilon_2 
+\epsilon_1^2+\epsilon_2^2}\over{(\gamma_{22}+\epsilon_2)^2+\epsilon_1^2}},
\eqno(15)
$$
and
$$
E_i={{\epsilon_1 (\gamma_{12}+\gamma_{22})^2}
\over{(\gamma_{22}+\epsilon_2)^2+\epsilon_1^2}},
\eqno(16)
$$
where $\gamma_{ij}=\rho_{ij}/ \rho$,
$\epsilon_1=(\gamma_{12}+\gamma_{22})
(f_c/f)\,F_r(\kappa)=(\gamma_{12}+\gamma_{22})
(f_c/f)\,F_r(\delta \sqrt{f/f_c})$,
$\epsilon_2=(\gamma_{12}+\gamma_{22})
(f_c/f)\,F_i(\kappa)=(\gamma_{12}+\gamma_{22})
(f_c/f)\,F_i(\delta \sqrt{f/f_c})$.
The function $F(\kappa)$ was written here more conveniently 
as a function of frequency $f$ , i.e.
$F(\kappa)=F(\delta \sqrt{f/f_c})$ \cite{biot},
where $\delta$ is a factor dependent on pore
geometry. For the hollow cylinder-like pores, 
$\delta=\sqrt{8}$ \cite{biot} and we use this value
throughout the paper. $f_c$ is the critical
frequency above which the Poiseuille flow breaks down,
and it equals  $b/(2 \pi \rho_2)$. 
Here $\rho_2$ denotes the product of porosity and
fluid mass density.

In order to obtain phase velocity and attenuation coefficient
of the rotational waves, we put $l=Re[l]+i Im[l]$. Thus, the phase
velocity is then $v_r= \chi / |Re[l]|$. Introducing
a reference velocity as $V_r=\sqrt{N/ \rho}$, we obtain the
dimensionless phase velocity as
$$
{{v_r}\over{V_r}}={{\sqrt{2}}\over{\left[ \sqrt{E_i^2+E_r^2}
+E_r\right]^{1/2}}}.
\eqno(17)
$$

To obtain the attenuation coefficient
of the rotational waves, we introduce
a reference length, $L_r$,  defined as
$L_r=V_r/(2 \pi f_c)$. The length $x_a$
represents the distance over which the
rotational wave amplitude is attenuated by a factor
of $1/e$. Therefore we can construct the dimensionless
attenuation coefficient as $L_r/x_a$,
$$
{{L_r}\over{x_a}}={{f}\over{f_c}}
{{\left[\sqrt{E_i^2+E_r^2} -E_r\right]^{1/2}}\over{\sqrt{2}}}.
\eqno(18)
$$

The equations describing dynamics of the dilatational waves are \cite{biot}
$$
\nabla^2(Pe+Q \epsilon)=
{{\partial^2}\over{\partial t^2}}(\rho_{11} e + \rho_{12} \epsilon)
+b F(\kappa){{\partial}\over{\partial t}}(e - \epsilon),
\eqno(19)
$$
$$
\nabla^2(Qe+R \epsilon)=
{{\partial^2}\over{\partial t^2}}(\rho_{12} e + \rho_{22} \epsilon)
-b F(\kappa){{\partial}\over{\partial t}}(e - \epsilon),
\eqno(20)
$$
where, $P,Q$ and $R$ are the elastic coefficients, $e=\mathrm{div} \;
\vec u$ and 
$\epsilon=\mathrm{div} \; \vec U$ are the divergence of solid 
and fluid displacements.
Again, substitution of a plane dilatational wave of the form
$$
e=C_1 e^{i(l x + \chi t)}, \;\;\;
\epsilon=C_2 e^{i(l x + \chi t)},
\eqno(21)
$$
into Eqs.(19) and (20) allows us to obtain a characteristic equation
$$
(z-z_1)(z-z_2)+ i M(z-1)=0,
\eqno(22)
$$
where $z=l^2 V_c^2/ \chi^2$, $V_c^2=(P+R+2Q)/ \rho$ represents
the velocity of a dilatational wave when the
relative motion between fluid and solid is absent,
$z_{1,2}=V_c^2/V_{1,2}^2$ with $V_{1,2}$ being the velocities
of the purely elastic waves with subscripts 1,2 referring to
the two roots of Eq.(22), and finally 
$M=(\epsilon_1 + i \epsilon_2)/(\sigma_{11} \sigma_{22}- \sigma_{12}^2)$
with $\sigma_{11}=P/(P+R+2Q)$, $\sigma_{22}=R/(P+R+2Q)$ and 
$\sigma_{12}=Q/(P+R+2Q)$.

Eq.(22) has two complex roots $z_I$ and $z_{II}$.
Phase velocities of the two kinds of dilatational waves
can be defined as 
$$
{{v_I}\over{V_c}}={{1}\over{Re[\sqrt{z_I}]}},
\;\;\; 
{{v_{II}}\over{V_c}}={{1}\over{Re[\sqrt{z_{II}}]}},
\eqno(23)
$$
while the corresponding attenuation coefficients can be
also introduced as
$$
{{L_c}\over{x_I}}={Im[\sqrt{z_I}]}{{f}\over{f_c}},
\;\;\; 
{{L_c}\over{x_{II}}}={Im[\sqrt{z_{II}}]}{{f}\over{f_c}}.
\eqno(24)
$$

\section{Numerical results}

In order to investigate the novelties brought about
into classical Biot's theory of propagation
of elastic waves in porous medium \cite{biot} by the
inclusion of boundary slip, we have
studied the full parameter space of the problem.

In all forthcoming results, we calculate phase velocities and
attenuation coefficients for the case 1 from Table I
taken from Ref. \cite{biot}, which is $\sigma_{11}=0.610$, 
$\sigma_{22}=0.305$, $\sigma_{12}=0.043$, $\gamma_{11}=0.500$,
$\gamma_{22}=0.500$, $\gamma_{12}=0$, $z_{1}=0.812$, and 
$z_{2}=1.674$.

We calculated normalized phase velocity
of the plane rotational waves, $v_r/V_r$, 
and the attenuation coefficient $L_r/x_{a}$ using 
our more general expression for $F(\kappa)$ 
(which takes into account non-zero boundary slip)
given by Eq.(7).

In Fig.~3 we plot phase velocity 
$v_r/V_r$ as a function of frequency for the three
cases: the solid  curve corresponds to $\xi=0$
(no boundary slip), while long-dashed and short-dashed curves correspond to
$\xi=1.0$ and $\xi=1.5$ respectively (we have used these
large values in order to emphasize the effect of the boundary slip).
Note that the $\xi=0$ case perfectly reproduces
the curve 1 in Fig. 5 from Ref.\cite{biot}. For $\xi=1.0$ and $1.5$
we notice a deviation from the classic 
behavior in the form of an decrease of phase
velocity in the domain of intermediate frequencies.

Fig.~4 shows the attenuation coefficient  
$L_r/x_a$ of the rotational wave
 as a function of frequency for the three
values of $\xi$: the solid  curve corresponds to $\xi=0$
(no boundary slip), while long-dashed and short-dashed curves represent 
the cases $\xi=1.0$ and $\xi=1.5$ respectively.
Note that $\xi=0$ case  coincides with
curve 1 in Fig. 6 from Ref.\cite{biot}. 
We observe that, in the intermediate frequency range,
with increase of boundary slip our
model yields substantially {\it higher}
values of the attenuation coefficient
than in the classic
no-slip case, which is in qualitative agreement with 
the experimental data \cite{gist}, \cite{mochizuki}.

We also calculated normalized phase velocities
of the plane dilatational waves, $v_I/V_c$ and 
$v_{II}/V_c$, and the attenuation coefficients $L_c/x_{I}$
and $L_c/x_{II}$   using 
our more general expression for $F(\kappa)$ 
given by Eq. (7).

In Figs.~5 and 6 are similar to Fig.~3 except that now 
we plot phase velocities
$v_I/V_c$ and $v_{II}/V_c$ as a function of frequency
for $\xi=0,\,0.05,\,0.1$. 
Note that the solid curves on the both graphs reproduce
curves 1 in Figs. 11 and 12 from Ref. \cite{biot}. 
We gather from the graph that introduction of the non-zero
boundary slip leads to decrease of $v_I/V_c$,
while $v_{II}/V_c$ increases in the 
in the domain of intermediate frequencies
as boundary slip, $\xi$, increases.

In Figs.~7 and 8 we plot the attenuation coefficients  
$L_c/x_I$ and $L_c/x_{II}$ in a similar manner as in Fig.~4,
but now for $\xi=0,\,0.05,\,0.1$.
Again we observe that the $\xi=0$ case  reproduces
curves 1 in Figs. 13 and 14 from Ref.\cite{biot}. 
We gather from Figs.~7 and 8 that introduction of the
non-zero boundary slip yields {\it higher}
than in the no boundary slip case values of the
attenuation coefficients for both types
of the dilatational waves in the domain of intermediate frequencies,
which, again,  is in qualitative agreement with 
the experimental results \cite{gist}, \cite{mochizuki}.

It is worthwhile to note that for no-zero $\xi$ the asymptotic behavior
of the elastic waves in the $f/f_c \to \infty$  as well as
 $f/f_c \to 0$ limit is identical to the classic
behavior established by Biot \cite{biot}.
This is a consequence of the assumptions of our
phenomenological model, which is based on
robust physical arguments.

\section{discussion}

In this paper we have studied the 
effect of introduction of the boundary slip 
in the theory of propagation 
of elastic waves in a fluid-saturated
porous solid originally formulated by Biot \cite{biot}.
Biot's theory does not account for boundary slip effect, however,
the boundary slip becomes important when the length scale over
which the fluid velocity changes approaches the slip length,
i.e. when  the fluid is highly confined, 
for instance, fluid flow through porous rock or 
blood vessel capillaries.
In the light of recent 
convincing experimental evidence of a boundary
slip in a Newtonian liquid  \cite{zg},
it is necessary to take into account this effect
 into the Biot's theory {\it where appropriate}.
We have studied the 
effect of introduction of boundary slip upon
the function $F(\kappa)$
 that measures the deviation from
Poiseuille flow friction as a function of frequency 
parameter $\kappa$.
This function crucially enters Biot's equations
which describe dynamics of fluid-saturated porous solid.
Therefore, a revision of Biot's theory was needed
in order to incorporate boundary slip effect into the
all measurable predictions of this theory such as
phase velocities
and attenuation coefficients of the both rotational 
and dilatational waves. We have performed such analysis, and
in summary, we found that the introduction of the
non-zero boundary slip into the Biot's theory of 
propagation of the elastic
waves in a fluid-saturated porous solid
results in
\begin{itemize}
\item the decrease,  
as compared to the no-slip limiting case, 
of the phase velocities of both rotational waves ($v_r/V_r$)
and dilatational wave of the first kind ($v_{I}/V_c$)
in the domain of intermediate frequencies. 
On contrary, the phase velocity of the
dilatational wave of the second kind ($v_{II}/V_c$)
experiences an increase as compared to the no-slip limiting case
in the domain of intermediate frequencies.

\item in the domain of intermediate frequencies
the attenuation coefficients of both the rotational ($L_r/x_a$) 
and dilatational waves ($L_c/x_{I}$ and $L_c/x_{II}$)
are {\it increased}  as compared to the no-slip limiting case
as the boundary slip increases, which is in qualitative agreement with 
the experimental data \cite{gist}, \cite{mochizuki}.

\item behavior of all physical
quantities which describe the elastic waves in the 
asymptotic limits of both small and large frequencies is 
{\it not} affected by the introduction of the 
non-zero boundary slip. The deviation occurs only
in the domain of intermediate frequencies, as prescribed
by our phenomenological model.
\end{itemize}

The investigation of properties of  elastic
waves is  important for a number of
applications. The knowledge of phase velocities and
attenuation coefficients of  elastic waves 
is necessary, for example,  to guide the oil-field exploration 
applications, acoustic stimulation of oil
producing fields (in order to increase the amount of recovered
residual oil), and the acoustic 
clean up of contaminated aquifers \cite{ore1,ore2,ore3,ore4}.
Therefore, our results would be useful for various
applications in oil production as well as in ecology.

From the recent experimental results of Ref.\cite{zg}
we gathered that there are physical situations were
the no-slip boundary condition becomes invalid.
We have formulated a {\it phenomenological} model of
elastic waves in the fluid-saturated porous medium
based on Biot's linear theory and certain physically
justified assumptions on the variation of boundary
slip velocity with frequency, $\bar U_s(\kappa)$.
Since, there are no experimental measurements of
$\bar U_s(\kappa)$, for a cylindrical tube, on which
"Biot-like" theory relies, there is a certain freedom
of choice, which could be used to obtain a better
fit of experimental data with the theory {\it in cases where
classic Biot's theory fails to do so}.
If fact, our model predicts {\it higher than the Biot's theory 
values of attenuation
coefficients of the both rotational 
and dilatational waves} in the intermediate frequency domain, 
which is in qualitative agreement with 
the experimental data \cite{gist}, \cite{mochizuki}.
Therefore, the introduction of the boundary slip yields three-fold benefits:
\begin{itemize}
\item Better agreement of theory with an experimental
data, since, the parametric space of the model is larger
(includes effects of boundary slip).
\item Possibility to identify types of porous medium and physical
situations where boundary slip is important.
\item Constrain model parameters that are related to the
boundary slip.
\end{itemize}

We would like to close this paper with the following two remarks:
First, the reported slip length varies from tens of nano-meters \cite{craig}
to a few microns \cite{zg}, depending on a type of a liquid and differences
in the experimental set up. The latter experimental work
clearly demonstrates the applicability of our
model to the elastic wave phenomena in a usual porous media 
found in Nature.
Second, in this work we {\it postulated} a phenomenological expression
for the boundary slip velocity based on some physically sound arguments
(in fact, similar, successful, approach has been used by the 
authors of Ref.\cite{tn01} in the context of solar physics). 
However, perhaps it may be possible to derive an expression based on
more general "first principles", which should be a subject of
a future work.

\newpage
\centerline{Figure Captions}
Fig.~1: Behavior of $Re [\bar U_s(\kappa)]=Im [\bar U_s(\kappa)]$
as function of frequency parameter, $\kappa$, for $\xi=1$, $A=25$.

Fig.~2: Behavior of $F_r (\kappa)$ (thick curves) $F_i (\kappa)$
(thin curves) as function of frequency parameter, $\kappa$, according to Eq.(7).
Solid curves correspond to the case when $\xi=0$ (no boundary slip),
while long-dashed and short-dashed curves correspond to
$\xi=0.05$ and $\xi=0.1$ respectively.

Fig.~3: Behavior of dimensionless, normalized phase velocity of
the rotational wave, $v_r/V_r$, 
as a function of frequency. Solid curve corresponds to the 
case when $\xi=0$ (no boundary slip),
while long-dashed and short-dashed curves correspond to
$\xi=1.0$ and $\xi=1.5$ respectively.

Fig.~4: Behavior of dimensionless, normalized attenuation coefficient of
the rotational wave, $L_r/x_a$, 
as a function of frequency. Solid curve corresponds to the 
case when $\xi=0$ (no boundary slip),
while long-dashed and short-dashed curves correspond to
$\xi=1.0$ and $\xi=1.5$ respectively.

Fig.~5: Behavior of dimensionless, normalized phase velocity of
the dilatational wave, $v_I/V_c$,
as a function of frequency. 
Solid curve corresponds to the 
case when $\xi=0$ (no boundary slip),
while long-dashed and short-dashed curves correspond to
$\xi=0.05$ and $\xi=0.1$ respectively.

Fig.~6: The same as in Fig.~5 except for the curves are for 
$v_{II}/V_c$.

Fig.~7: Behavior of dimensionless, normalized attenuation coefficient of
the dilatational wave, $L_c/x_I$, 
as a function of frequency. Solid curve corresponds to the 
case when $\xi=0$ (no boundary slip),
while long-dashed and short-dashed curves correspond to
$\xi=0.05$ and $\xi=0.1$ respectively.

Fig.~8: The same as in Fig.~7 except for the curves are for 
$L_c/x_{II}$.
%\newpage
%\underline{Short Title:} Phenomenological model of non-zero boundary slip
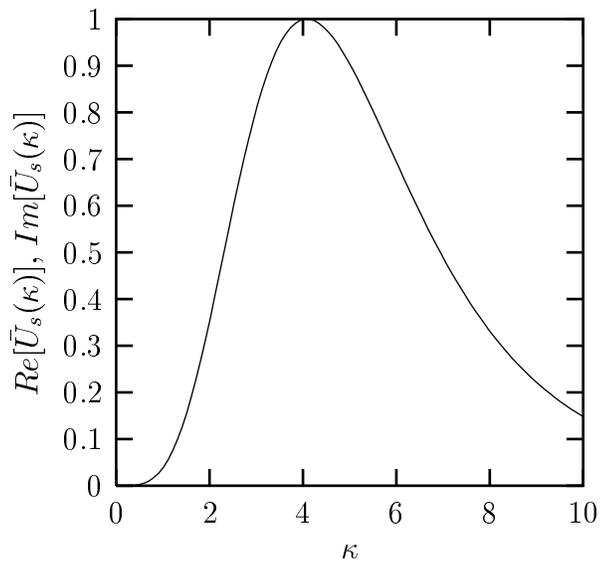
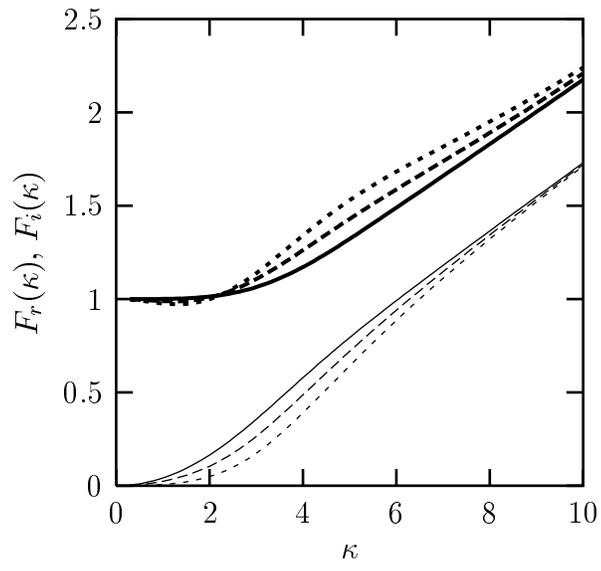
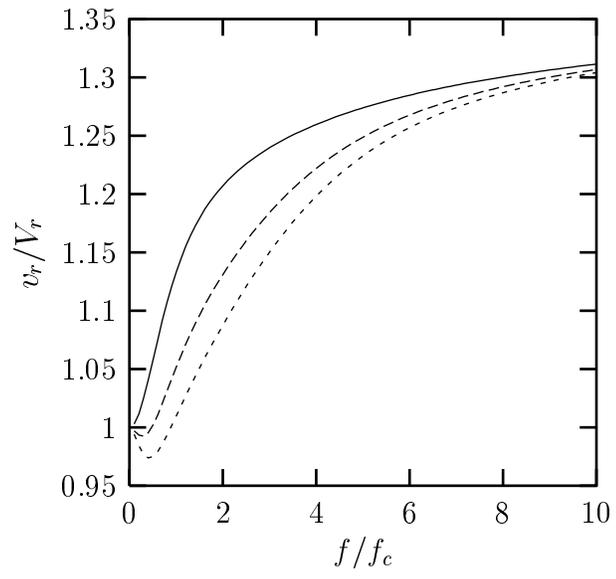
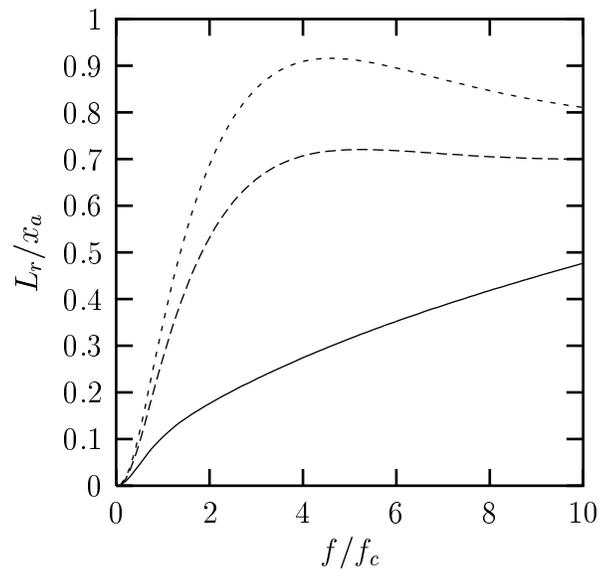
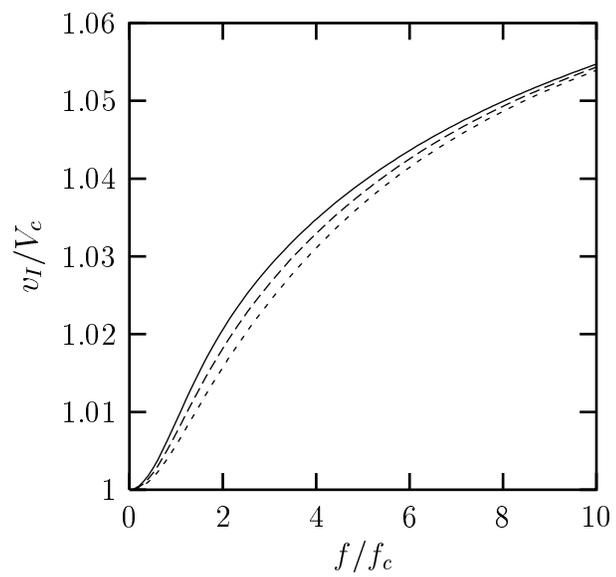
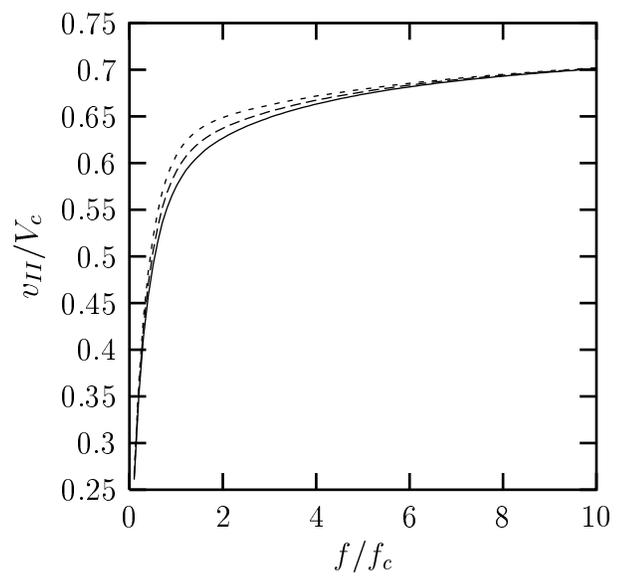
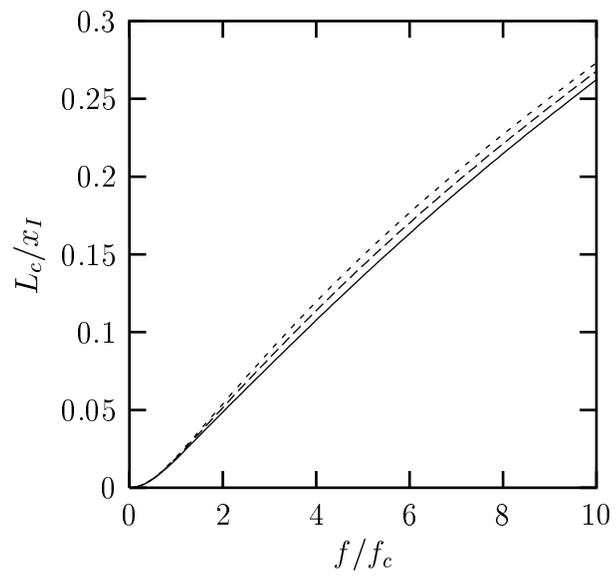
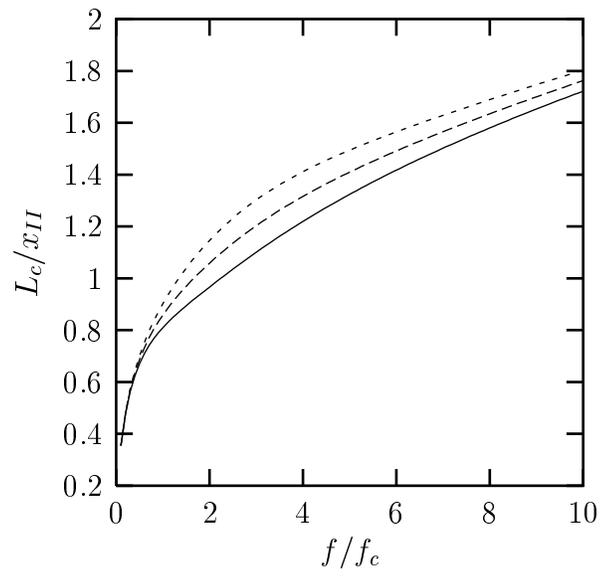
\end{document}